\documentclass[a4paper,11pt]{article}
\usepackage{pos}

\def\lamA{\lambda_{\scriptscriptstyle{\text{\rm A}}}}        
\def\fir{{\scriptscriptstyle{\text{\rm IR}}}}                 
\def\fuv{{\scriptscriptstyle{\text{\rm UV}}}}              
\def\cro{{\scriptscriptstyle{\text{\rm A}}}}                 
\def\efN{\mathscr{N}}                                        
\def\efNm{\efN_\star}                                        

\title{The Infrared Phase of QCD and Anderson Localization}

\author*[a,b,c]{Ivan Horv\'ath}

\affiliation[a]{Nuclear Physics Institute CAS,\\
  Hlavn\'i 130, 25068 \v{R}e\v{z} (Prague), Czech Republic}

\affiliation[b]{Department of Physics and Astronomy, University of Kentucky,\\
506 Library Drive, Lexington KY 40506, USA}

\affiliation[c]{Department of Physics, The George Washington University,\\
725 21st St. NW, Washington DC 20052, USA}

\emailAdd{horvath@ujf.cas.cz}

\abstract{When Anderson localization entered the QCD landscape, it was almost 
immediately thought about in connection with thermal phases, namely as a factor 
in the chiral transition. However, recent developments revealed an additional 
structure that made Anderson-like features central to the genesis of the entirely 
new thermal phase: the IR phase. I will explain these developments.}

\FullConference{The XVIth Quark Confinement and the Hadron Spectrum Conference (QCHSC24)\\
 19-24 August, 2024\\
 Cairns Convention Centre, Cairns, Queensland, Australia\\}

\begin{document}
\maketitle

\section{Introduction} 
\label{sec:intro}
In this talk I will describe what is shaping up as a new case of fruitful 
cross-fertilization between elementary particle and condensed matter physics. 
On the former side it involves a recent unexpected finding of a new thermal 
regime in 
QCD~\cite{Alexandru:2019gdm, Alexandru:2015fxa, Alexandru:2021pap, 
Alexandru:2021xoi}, characterized by proliferation of deeply infrared 
($0 \!\lesssim\! \lambda \!\ll\! T$) Dirac modes, by decoupling of 
the infrared physics (IR-bulk separation), and by scale invariant glue in 
the ensuing IR component.
This new regime is known as the {\em IR phase} and is interesting not only 
due to the novel nature of its thermal state but also due to the fact that 
the associated change may be a true phase transition occurring at temperature 
$T_\fir$ ($200 \,\text{MeV} \!< T_\fir \!< 230\,$MeV) just above the known 
range of crossover temperatures.

On the other side of the above relationship is the time-honored phenomenon 
of Anderson localization, namely the exponential pinning of a quantum particle 
by virtue of spatial disorder~\cite{Anderson:1958a, doi:10.1142/7663}. This
effect had a deep impact on condensed matter physics and various applied 
areas but its potential role at the fundamental level, such as in the Standard 
Model of elementary particles, was barely considered in the literature. 
A fruitful remark~\cite{Halasz:1995vd} suggested that, at sufficiently high
temperatures, thermal fluctuations in the strong sector (QCD) may induce 
localization of quark modes in low-lying parts of Dirac spectra. 
The associated loss of IR quark mobility then led to the suggestion that QCD 
chiral transition (massless limit), and by proxy the transition in real-world QCD, 
could be viewed as analogous to metal-to-insulator transition of Anderson
type~\cite{GarciaGarcia:2006gr}. Such considerations 
gained some traction when localization in hot QCD started to be investigated 
more systematically and the Anderson-like mobility edge $\lamA \!>\! 0$ was 
identified\footnote{Note that in this continuum notation, where non-zero 
Dirac eigenvalues come in conjugate pairs $\pm i \lambda$, we frequently
only consider the upper branch $\lambda \!\ge\! 0 $.} at sufficiently high 
temperatures~\cite{Kovacs:2010wx, Giordano:2013taa}.

But contradictions soon appeared as well~\cite{Alexandru:2014paa, 
Alexandru:2015fxa}. To that end, note that the ``insulating'' nature of hot 
thermal state in metal-to-insulator scenario signifies the absence of 
long-distance ($\gtrapprox \,$1/T) physics. This was assumed to arise due 
to the infrared (IR) modes being localized at shorter scales, and due 
to the expected depletion of Dirac spectrum in this range. 
But the strong proliferation of deep-IR modes at hight temperatures
contrary to expectations~\cite{Alexandru:2015fxa}, and their chiral 
polarization properties~\cite{Alexandru:2014paa}, 
suggested that IR physics is actually present, raising the possibility that metal-to-insulator 
scenario may not reflect some key aspects of QCD reality. 

This developing discord became sharp once the evidence for IR phase 
was presented and its defining features were laid
down~\cite{Alexandru:2019gdm}. In this original work it was 
already proposed that the mobility edge $\lamA \! > \! 0$ is crucial 
for IR-bulk separation, and that non-analyticity introduced by it shields 
the IR from renormalization-group running in line with the observed 
IR scale invariance. But the nature of the emerged IR physics 
remained foggy. The crucial step in this regard was taken in 
Ref.~\cite{Alexandru:2021pap} which studied effective spatial 
dimensions of IR modes and, apart from intriguing non-analytic 
behavior (in $\lambda$) suggesting topology at play, found that 
point $\lambda_\fir \!\equiv\! 0$ behaves in certain regards similarly 
to the mobility edge $\lamA$. Ref.~\cite{Alexandru:2021xoi} then 
provided evidence that $\lambda_\fir$ is indeed a new Anderson-like 
mobility edge, and formulated the {\em metal-to-critical} scenario 
of transition to IR phase based on both edges.

\begin{figure}
	\centering
	\vskip -0.16in	
	\includegraphics[width=0.58\linewidth]{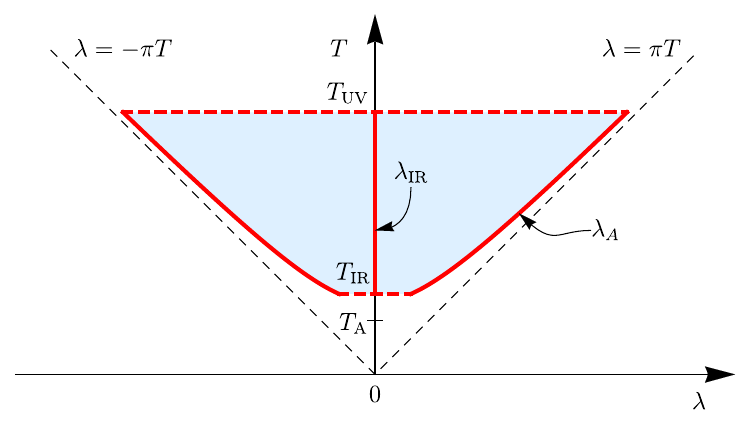}
	\vskip -0.06in	
	\caption{\label{fig:dir_phase_diagram} Phase diagram for Anderson-like 
	properties in the Dirac spectrum of thermal QCD~\cite{Alexandru:2021xoi}. 
	See the explanation in the text. Temperature $T_\cro$ is the crossover 
	point for Dirac spectral properties~\cite{Alexandru:2019gdm} and thin 
	dashed lines represent the Matsubara scales.}  
	\vskip -0.16in
\end{figure}

The metal-to-critical scenario is conveyed by the phase diagram for
Anderson-like properties of thermal QCD~ \cite{Alexandru:2021xoi}, 
shown in Fig.~\ref{fig:dir_phase_diagram}. 
Here the thick red lines in $\lambda \!-\! T$ 
plane represent the Anderson-like critical points (mobility edges). 
Dashing at the horizontal segment $T \!=\! T_\fir$ signifies that it may 
be of zero length (single point), and dashing at $T \!=\! T_\fuv$ 
conveys that $T_\fuv$ may be infinite. 
In IR phase, namely for $T_\fir < T < T_\fuv$, there are critical points 
$\pm \lamA$ and $\lambda_\fir$ in the Dirac spectrum, yielding 
the modes in the range $(-\lamA, \lambda_\fir) \cup (\lambda_\fir, \lamA)$ 
(the blue region) localized with varying localization lengths that diverge at 
$\pm \lamA$ and $\lambda_\fir$. The chief tenet of the relationship 
between the thermal state of QCD in IR phase and its Anderson-like 
representation (metal-to-critical picture) is that the above constellation of 
critical points captures the unusual features found in 
the former: the non-analyticity at $\lamA$ facilitates the IR-bulk separation 
and shields the IR from renormalization-group running, while the criticality 
at $\lambda_\fir$ determines the nature of its long-range physics.

In this talk I will describe in some detail the developments that led to the current 
notion of IR phase and emphasize, in particular, the role of Anderson 
localization ideas in that process.


\section{IR Phase} 
Initial key steps toward the notion of IR phase were taken in 
Ref.~\cite{Alexandru:2015fxa}. 
The authors argued that a qualitatively new regime 
characterized by anomalously strong accumulation of IR Dirac modes exists in 
SU(3) gauge theories with fundamental quarks. They showed that this accumulation, 
observed in lattice-regularized theory, persists into the continuum limit and suggested 
that the ensuing IR degrees of freedom (dof's) decouple and defy confinement 
in this ``partially deconfined''~phase.
    
The proliferation of IR dof's remained the chief qualitative underpinning 
of the new regime but the concept of IR phase arose later with 
the revelation~\cite{Alexandru:2019gdm} that the effect is quantitatively 
expressed as a power-law (negative power) IR behavior of Dirac spectral 
density\footnote{Recall that Dirac spectral density $\rho(\lambda)$ of the theory 
is the average number of Dirac eigenmodes per unit 4-volume and unit spectral 
interval in the infinitesimal neighborhood of $\lambda$.}.
In fact, at least in thermal cases, the near-pure power $p\!=\!-1+\delta$ with very 
small $\delta \ge 0$ was found.
All basic tenets of IR phase were formulated in that work, and SU(3) gauge 
theories with fundamental quarks were classified into three types (phases) based 
on a degree of their Dirac mode accumulation 
in deep IR, namely~\cite{Alexandru:2019gdm} 
\begin{equation}
     \rho(\lambda)   \propto   \lambda^p   \; , \;    
     \lambda \to 0    \quad\;\, \Rightarrow \quad\;\,
     \text{phase}  \;=\;\,
     \text{B}  \;\;  \text{if}  \;\;  p=0 
     \;\;\; , \;\;\;\text{IR}   \;\;  \text{if} \;\;  p<0  
     \;\;\; , \;\;\;\text{UV}  \;\;  \text{if} \;\;  p>0 
     \;\;\;      
     \label{eq:010}
\end{equation}
The names of phases are derived from the proposed relationship of IR glue 
to scale invariance:  B is a shorthand for IR-Broken and corresponds to 
hadronic phase such as in real-world QCD at 
low temperatures.\footnote{Note that the logarithmic IR behaviors, such as 
one predicted in low-temperature QCD by chiral perturbation 
theory~\cite{Osborn:1998qb}, entail $p \!=\! 0$ and belong to B phase.} 
IR refers to the IR-Symmetric phase which is synonymous to the IR phase and 
for which a possible connection to near-perfect fluid medium observed at 
RHIC and LHC was raised~\cite{Alexandru:2019gdm}. UV dubs ``IR-Trivial'' 
since IR degrees of freedom are power-law suppressed. This regime would 
correspond to weakly-coupled quark-gluon plasma. Note that the nominal 
expectation is that $\rho(\lambda) \propto \lambda^3$ in UV 
($\lambda \to \infty$) for all three phases due to asymptotic freedom. 

\begin{figure}
	\centering  
	\vskip -0.08in	      
	\includegraphics[width=0.86\linewidth]{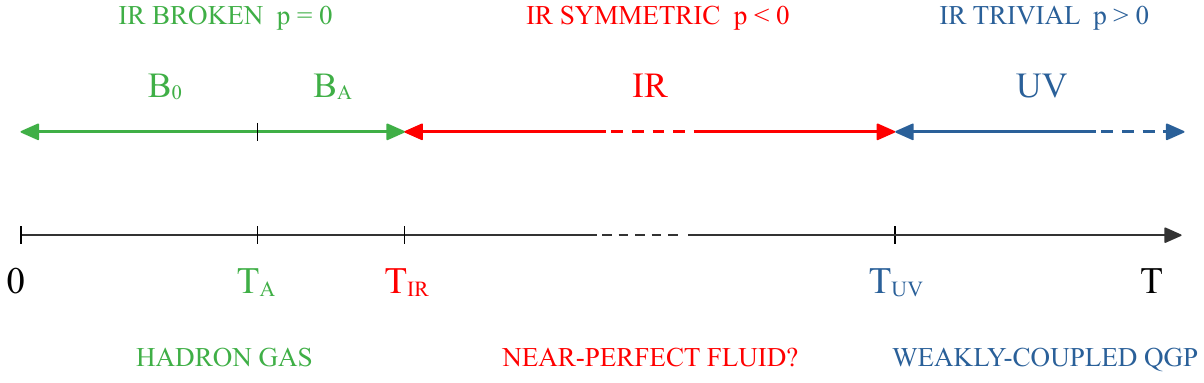}
	\caption{\label{fig:phases} Types of phases in SU(3) gauge theories 
	with fundamental quarks based on the abundance of deep-IR degrees 
	of freedom and IR scale invariance. See the discussion in the text.}  
	\vskip -0.14in
\end{figure}

The above is schematically represented by Fig.~\ref{fig:phases}. The order 
from left-to-right follows the case when transitions are induced by increasing 
temperature. Temperature $T_\cro$ marks a generic crossover in Dirac 
spectral properties. This transition is thus on the same footing as the chiral 
or other crossovers. Temperatures $T_\fir$ and $T_\fuv$ mark transitions 
to IR and UV phases respectively. 

\begin{figure}[b]
	\vskip -0.1in     
	\centering   
	\includegraphics[width=0.42\linewidth]{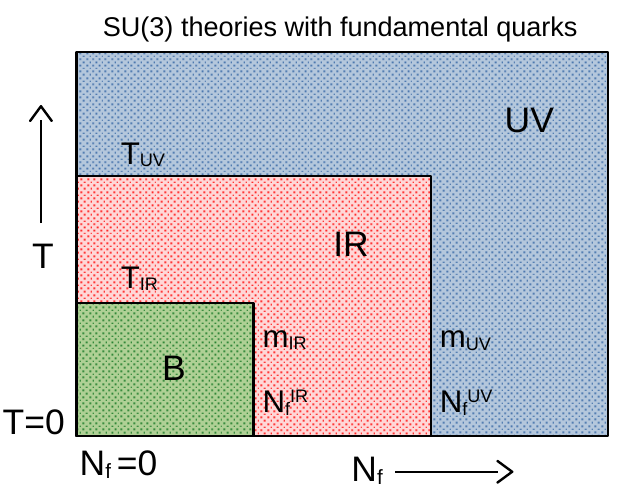}
	\hskip 0.35in
	\includegraphics[width=0.42\linewidth]{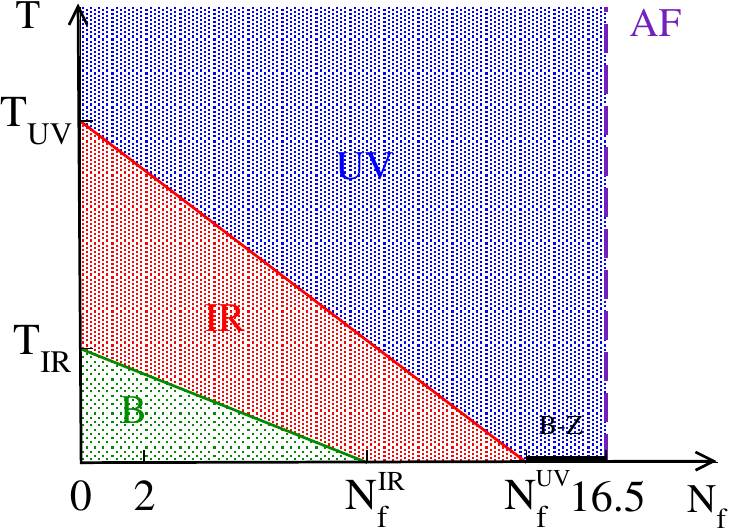}
	\caption{\label{fig:fullSU3} Left: schematic phase diagram of SU(3) 
	gauge theories with fundamental quarks (set ${\cal T}$) based on 
	the abundance of deep-IR degrees of freedom and IR scale invariance. 
	Right: the case of near-massless quarks. The asymptotic freedom (AF) 
	boundary is at $N_f \!=\! 16.5$. See the discussion in the text.}  
	\vskip -0.1in
\end{figure}

The phase structure in the full set ${\cal T}$ of SU(3) gauge theories with 
fundamental quarks is shown schematically in Fig.~\ref{fig:fullSU3} 
(left)~\cite{Alexandru:2015fxa}. ${\cal T}$ is a multidimensional 
space ($T$, number of flavors $N_f$, individual quark masses) and only 
$T$ and $N_f$ have an explicit representation in Fig.~\ref{fig:fullSU3} (left). 
The accumulated knowledge endowing the scheme with information is that 
the phase sequence $B \,\rightarrow\, IR \,\rightarrow\, UV$ (or any part of it) 
occurs not only via increasing the temperature, but also via increasing 
the number of flavors and via {\em decreasing} the individual quark masses $m_i$~\cite{Alexandru:2019gdm}. Important case is when IR phase arises 
from B phase at sufficiently large $N_f$ by lowering the quark masses alone. 
In the degenerate near-massless case this situation is schematically shown 
in Fig.~\ref{fig:fullSU3} (right), which for $T \rightarrow 0$ approaches what 
is known as the conformal window region~\cite{Banks:1981nn}. 
The corresponding transition parameters $N_f^\fir$ and $N_f^\fuv$ delimit 
its strongly-coupled 
part~\cite{Alexandru:2019gdm} while theories in the regime $(N_f^\fuv, 16.5)$ 
are expected to be governed by the weakly-coupled Banks-Zaks fixed 
point. Note that the straight-line phase boundaries in Fig.~\ref{fig:fullSU3} 
signify the lack of more detailed knowledge at present.   

\begin{figure}[t]
	\centering
	\vskip -0.16in	
	\includegraphics[width=0.55\linewidth]{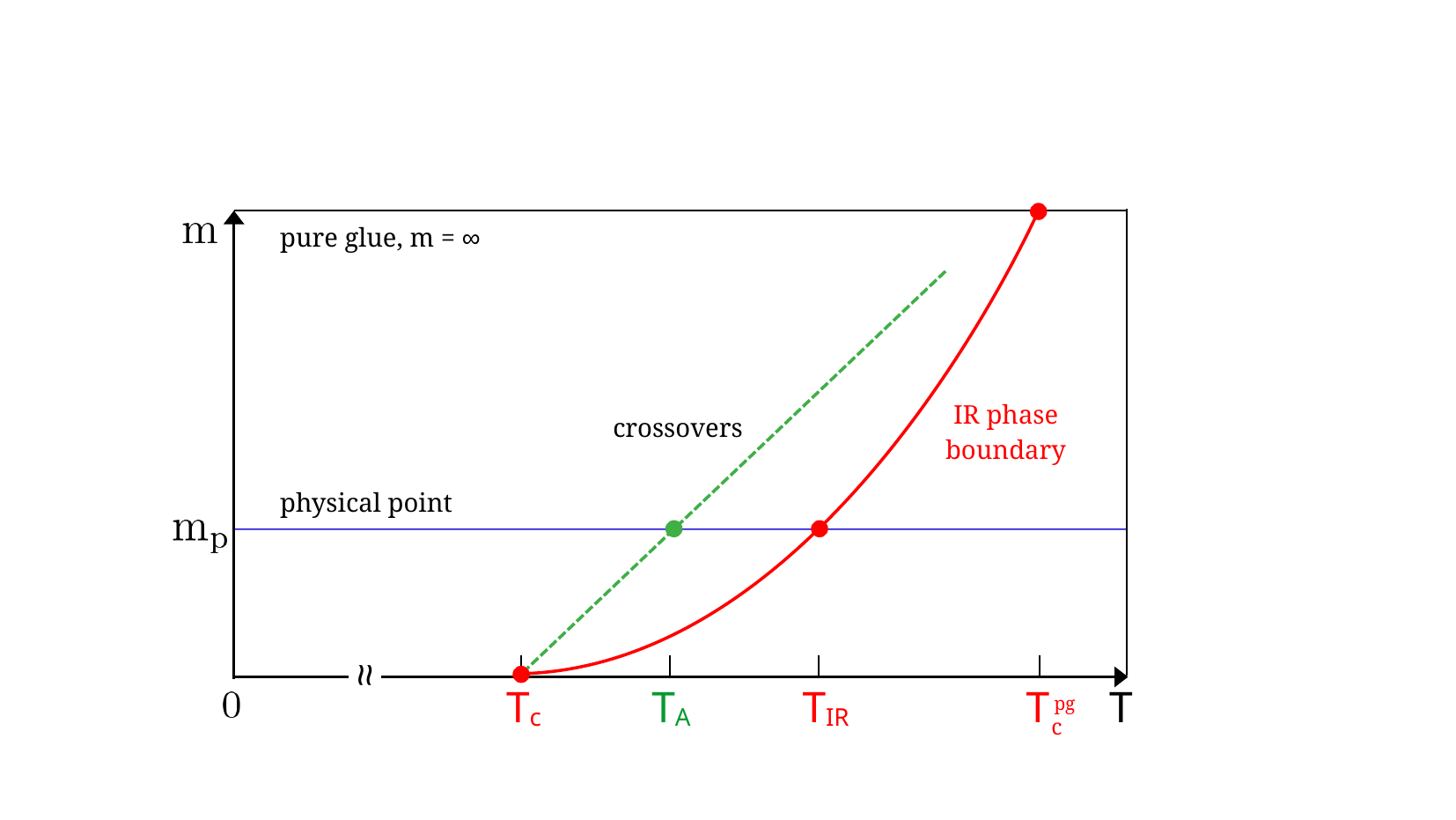}	
	\caption{\label{fig:two_flavors} The conjectured $N_f \!= 2$ (or $N_f \!= 2+1$) 
	thermal QCD phase diagram including the IR phase~\cite{Alexandru:2019gdm}. 
	See e.g. the talk by I.~Horv\'ath at FunQCD22 workshop for one of the explicit 
	mentions of this structure:
	\url{https://drive.google.com/file/d/1vZ0AY0WsZAfF9iV7-Br-E_2NiwaZzRGp/view}.}  
	\vskip -0.18in
\end{figure}

The important special case is the $T\!-\!m$ phase diagram for $N_f \!=\!2$ 
(mass-degenerate) flavors or for $N_f \!= 2\!+\!1$ flavors with heavy quark 
at the physical strange-quark mass. These are both very accurate
representations of  ``real-world QCD'' when light-quark mass $m$ is set 
to physical value. Varying $m$ probes the effects of light 
quarks and the approach to the chiral limit. The simplest scenario with
IR phase included is shown in Fig.~\ref{fig:two_flavors}. IR phase transitions
are traced by the red line starting at $m \!=\! 0$ with the chiral transition 
at $T_c$,  passing through $T_\fir$ at the physical point, and ending at 
the Polyakov-line transition of pure-glue theory ($m \!=\! \infty$) denoted 
as $T_c^{pg}$. Generic crossovers (e.g. the chiral one) are represented 
by the green dashed line, with $T_\cro$ denoting their temperature.

\subsection{Some Key Evidences}
\begin{figure}[t]
	\centering    
	\vskip -0.17in		    
	\includegraphics[width=0.96\linewidth]{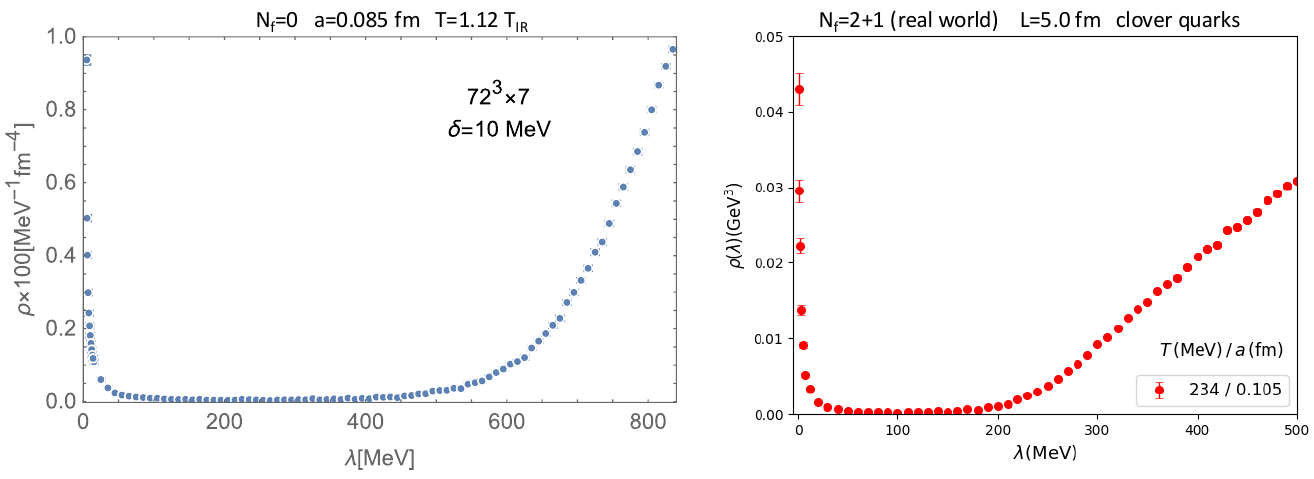}
	\vskip -0.06in	
	\caption{\label{fig:greatIR} Dirac spectral densities for pure-glue QCD 
	(left,~\cite{Alexandru:2024np}) and ``real-world" ($N_f \!= 2+1$) QCD 
	(right,~\cite{Meng:2023nxf}) in IR phase. The parameters of regularized 
	systems in question are specified in the plots.}
	\vskip -0.11in
\end{figure}

The power of numerical lattice QCD lies, among other things, in the fact that by 
virtue of fully realizing a regularized system, it can verify its conjectured  
features. By the same token, it can also discover qualitatively new aspects that 
were not even considered. IR phase is an example of this. Indeed, it was widely 
expected that thermal fluctuations universally suppress IR dof's and Dirac spectrum 
was generally assumed to become IR-depleted upon the ``QCD transition''.
Instead, Fig.~\ref{fig:greatIR} shows the state of the art results for spectral 
densities of the overlap Dirac operator in pure-glue (left) and real-world (right) QCD,
with an entirely different IR world clearly emerging at large 
volumes.

\begin{figure}[b] 
     \centering  
     \vskip -0.14in      
     \includegraphics[width=0.84\linewidth]{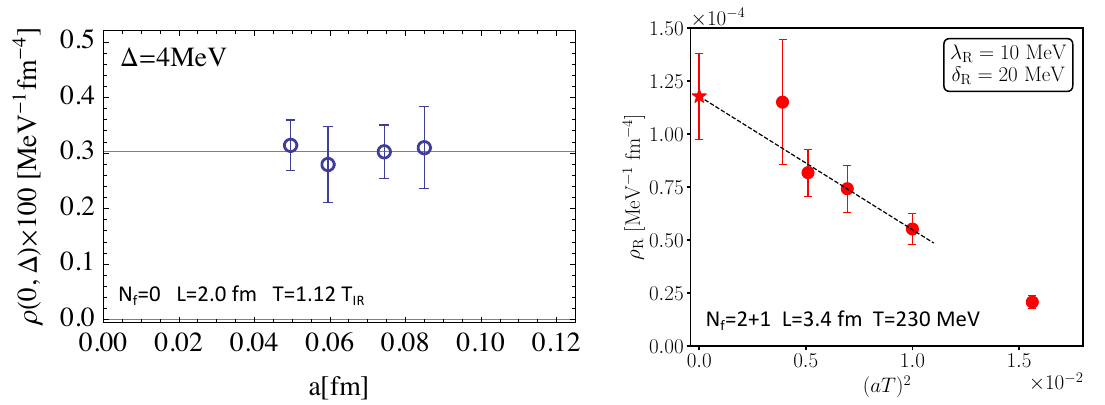}
     \vskip -0.02in           
     \caption{\label{fig:IR_scaling} Scaling of the strength of IR peak in 
     Dirac spectral densities for pure-glue QCD (left, 
     overlap,~\cite{Alexandru:2015fxa}) and 
     ``real-world" QCD (right,  staggered,~\cite{Alexandru:2024tel}). 
     The latter uses staggered dynamical lattice quarks.}
      \vskip -0.17in
\end{figure}

Two ingredients were of key relevance for the notion of IR phase to develop. 
The first one relates to the generic worry that lattice observations may be 
regularization artifacts that disappear in the continuum limit. 
This was resolved for pure-glue QCD in Ref.~\cite{Alexandru:2015fxa} 
and for ``real-world" QCD ($N_f \!= 2+1$ at physical point) in 
Ref.~\cite{Alexandru:2024tel}.  Fig.~\ref{fig:IR_scaling} shows the UV cutoff 
dependences of mode abundance in the inner core of the IR peak which
reliably extrapolate to non-zero values. 

\begin{figure}[t]
	\centering  
	\vskip -0.18in      
	\includegraphics[width=0.85\linewidth]{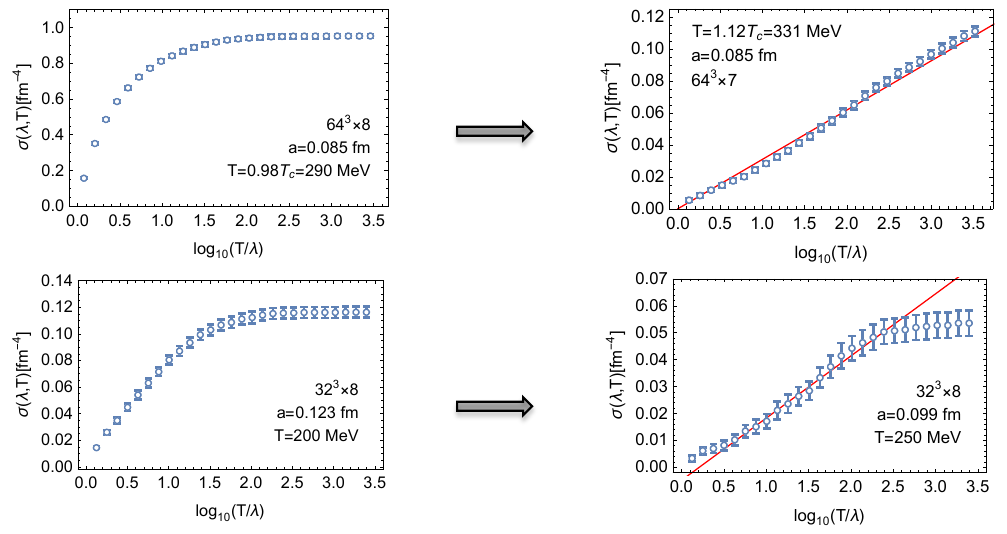}
	\vskip -0.08in      	
	\caption{\label{fig:transitions} Transitions to IR phase in pure-glue (top) and 
	``real-world'' QCD (bottom)~\cite{Alexandru:2019gdm}. See the explanation 
	in the text. Real-world QCD refers to $N_f \!=2+1$ at the physical point.}
	\vskip -0.18in
\end{figure}

The second key ingredient was the revelation that the unusual IR accumulation 
has a power-law character~\cite{Alexandru:2019gdm}. In particular, $\rho(\lambda)$ 
behaves in IR as a near-pure power $\lambda^{p}$ with $p\!=\! -1 + \delta$ 
and~$\delta \!>\! 0$ very small. 
This is shown in Fig.~\ref{fig:transitions} where the variables were chosen
so that pure $p\!=\!-1$ corresponds to a rising straight line. In particular, 
$\sigma(\lambda,T)$ is a cumulative density for the interval $(\lambda,T)$. 
Approaching deeper IR means moving to the right on the plots. Notice that in 
pure-glue case the drastic change occurs when changing $T\!=\!0.98 T_\fir$ 
to $T\!=\!1.12 T_\fir$ and the power law in IR phase persists for over three 
orders of magnitude in scale. Temperature $T_\fir$ coincides with $T_c$ 
of the Polyakov line transition~\cite{Alexandru:2019gdm}. 
The ``real-world" case behaves very similarly at $T \!=\! 250\,$MeV, although 
the range of power law is smaller due to smaller volumes.

\subsection{Why IR and Why Phase?}
The ``IR'' in IR phase arose from the qualitatively enhanced propensity of dof's to be 
arbitrarily infrared upon crossing into this new regime. The connection to IR fixed points 
associated with the conformal window offered itself naturally~\cite{Alexandru:2019gdm} 
although the details still need to be investigated. Most of the concluded features of 
the phase come from studies of thermal QCD with temperature being the transition-inducing
parameter. From the physics standpoint, this is arguably the most important setting 
to investigate in fact. At present, the best estimate for the value of $T_\fir$ in real-world
QCD comes from $N_f \!= 2+1$ studies at physical quark masses and
yield~\cite{Alexandru:2019gdm, Meng:2023nxf, Alexandru:2024tel}
\begin{equation}
    \text{200 MeV} \; < \; T_\fir \; < \; \text{230 MeV}
    \label{eq:020}
\end{equation}
Note that the well-known chiral crossover temperature is
$T_\cro \rightarrow T_\cro^c \!\approx\! 155\,$MeV~\cite{Aoki:2006we, Aoki:2009sc, 
HotQCD:2018pds}.

Several intriguing proposed features of the IR regime make it a phase in the conventional 
sense~\cite{Alexandru:2019gdm, Alexandru:2021pap,  Alexandru:2021xoi}. 
In particular, upon crossing into the IR phase the following occurs:

\begin{description}

   \item[{\em (i)}] {\bf IR-Bulk Separation.} The system becomes multicomponent with 
   the IR segment decoupling and becoming an autonomous subsystem (independent
   component).   
   
   \item[{\em (ii)}] {\bf IR Scale Invariance.}  Glue fields of the IR component 
   become scale invariant (at least asymptotically in IR). 
    
   \item[{\em (iii)}] {\bf Non-Analyticity.}  Non-analytic behavior in Dirac spectral
   properties (in $\lambda$-dependence) appears and translates into non-analytic 
   $T$-dependence of physical observables at the transition.
  
   \item[{\em (iv)}] {\bf Infinite Glue Screening Lengths.}  Gluon fields start 
   exhibiting long-range spatial correlations. 
                
\end{description}

In the following section we will describe how these features interrelate with 
the existence of Anderson-like transitions in high-temperature QCD.

\section{IR Phase and Anderson-like Localization}
\label{sec:AL_IR}

At the time the notion of IR phase was put forward in~\cite{Alexandru:2019gdm}, 
some of the arguments for justification of properties {\bf {\em (i)}}-{\bf {\em (iv)}} 
already utilized the existence of the Anderson-like mobility edge $\lamA \!>\! 0$ 
in Dirac spectra at high temperatures. When the metal-to-critical scenario
utilizing also the new mobility edge at $\lambda_\fir \!=\! 0$ was put 
forward~\cite{Alexandru:2021xoi, Alexandru:2021pap} (see Sec.~\ref{sec:intro}), 
essentially all proposed features of IR phase could be thought about under that 
umbrella. In this section we describe this ensuing relationship between 
the Anderson-like localization phenomena and thermal QCD. 
 
\begin{figure}[b]
	\centering  
	\vskip -0.14in      
	\includegraphics[width=0.75\linewidth]{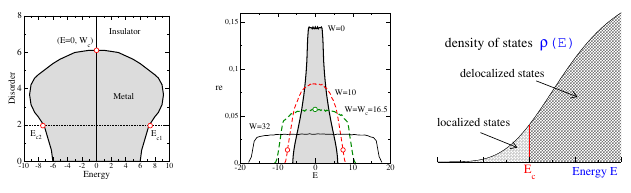}
	\vskip -0.06in      	
	\caption{\label{fig:AT} Generic schematics of Anderson transitions: phase diagram,
	density of states and the mobility edge. Courtesy of Peter Marko\v s and
	Ref.~\cite{Markos:2006}.}
	\vskip -0.18in
\end{figure}

\subsection{Anderson Mobility Edges}
Anderson localization is an old subject~\cite{Anderson:1958a} that became
quite vast in part due to the wealth of its applications~\cite{doi:10.1142/7663}. 
For present purposes we only need a cursory knowledge of the topic, mostly 
embodied in the abstract notion of Anderson transitions or the associated
critical points: the mobility edges. On one side of the transition are extended
(``conducting'') states while on the other side the exponentially localized 
(``insulating'') ones. Transition occurs due to spatial disorder (e.g. random 
potentials), and the criticality at the edge is conveyed, among other things, 
by the fact that the density-density (probability-probability) spatial correlation 
length in modes diverges. Schematics of these transitions are 
shown in Fig.~\ref{fig:AT}. For example, when approaching the mobility 
edge indicated by energy $E_{c1} \!>\! 0$ on the phase diagram (left plot) 
from the localized side $E \!>\! E_{c1}$, then the geometric scale $\ell$ 
(mode size, ``localization length'') associated with the mode varies as 
\begin{equation}
    \ell(E) \,\propto\, (E-E_{c1})^{-\nu} \,\propto \,  \xi(E) 
    \qquad\quad
    \nu > 0
    \label{eq:030}
\end{equation}
and is proportional to the correlation length $\xi$. Directly at criticality, 
the density-density correlation function becomes long-range namely  
$L^6 \langle p(r) p(r') \rangle \propto (| r-r' |/L)^{-\eta}$ in 3D 
(see e.g.~\cite{Evers_2008}). Here $p(r) = \psi^\dagger \psi (r)$ with 
$\psi$ the eigenstate.


\subsection{IR-Bulk Separation}
The idea of IR-Bulk separation became suggestive after the striking bimodal 
structure of spectral densities became the fact of the continuum 
limit~\cite{Alexandru:2015fxa}. However, that in itself doesn't guarantee that
the modes, and thus the physics itself, associated with spectral segments so 
separated are fully independent of one-another. One possibility that would 
make this very assuring is e.g. if $\rho(\lambda)$ had a region of full 
depletion ($\rho = 0$) between the two parts. While it is very possible that 
this does happen upon removing both the IR and UV cutoffs, it has not been 
verified in simulations yet. 

But it is also not necessary. Indeed, there may be a large class of other 
spectral non-analyticities separating IR and Bulk that indicate their 
decoupling. To that end, the original work~\cite{Alexandru:2019gdm} 
suggested that Anderson-like mobility edge 
$\lamA \!>\! 0$~\cite{Kovacs:2010wx, Giordano:2013taa} fits the bill in terms 
of providing for the needed non-analytic setup. Indeed, it is a critical point 
(internal ``phase transition'') within the Dirac substructure of QCD, and 
modes below and above $\lamA$ are thus not expected to talk to each other. 
In fact, if $\lamA$ is truly Anderson-like then spatial correlations across this 
edge explicitly conform to that~\cite{doi:10.1142/7663}. 
Thus, in the metal-to-critical picture of IR phase transition, IR-bulk 
separation is produced~via~$\lamA$.

\subsection{IR Scale Invariance and Infinite Glue Screening Lengths}
The chief initial motivation for IR scale invariance in IR 
phase~\cite{Alexandru:2019gdm} was the near-pure power-law behavior 
of $\rho(\lambda)$: in the spirit of the inverse scattering problem in 
quantum mechanics, this was suggested to be the result of scale 
invariance (in the statistical sense) of the underlying glue. Another 
motivation was that similar IR accumulation of Dirac modes also 
occurs in the vicinity of conformal window ($T\!=\!0$) as a result of 
lowering quark masses~\cite{Alexandru:2014paa, Alexandru:2014zna}. 
This then suggested the contiguous IR phase region in the theory space 
(Fig.~\ref{fig:fullSU3}) and made IR scale invariance natural in 
that sense.  

The metal-to-critical scenario of transition to IR phase associates 
the appearance of long-range correlations (power-law decays; infinite 
correlation lengths) with the formation of strictly IR Anderson-like 
mobility edge $\lamA \!=\! 0$ upon crossing 
$T_\fir$~\cite{Alexandru:2021xoi, Alexandru:2021pap}. 
Indeed, identifying the IR phase transition with the emergence of 
$\lamA$ means that critical correlations in deep-IR Dirac modes 
ensue upon entering the phase. This, in turn, translates into long-range 
correlations in gluonic composite fields which can be scale-decomposed 
in terms of these 
eigenmodes~\cite{Horvath:2006az, Horvath:2006md, Alexandru:2008fu}. 

Metal-to-critical scenario thus not only incorporates infinite glue screening 
lengths and (at least an asymptotic) IR scale invariance of glue in the IR 
phase, but also provides for a concrete and from the particle physics point 
of view highly unusual mechanism of their origin. 

\begin{figure}[t]
	\centering  
	\vskip -0.16in      
	\includegraphics[width=0.56\linewidth]{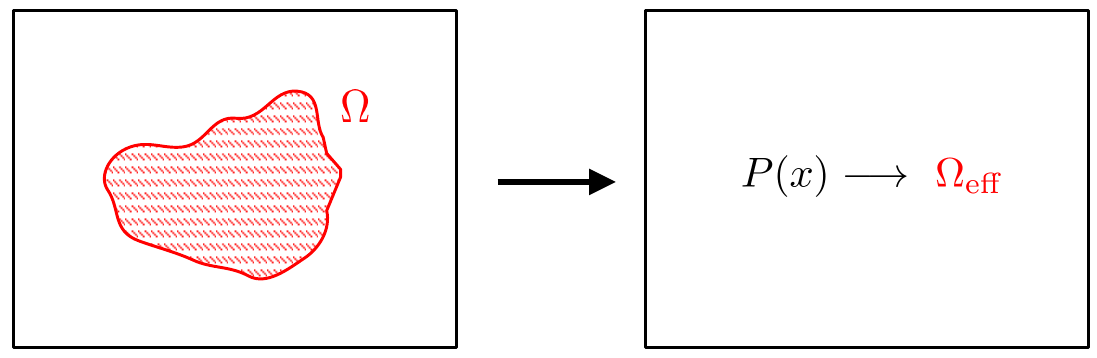}
	\hskip 0.70in		
	\includegraphics[width=0.254\linewidth]{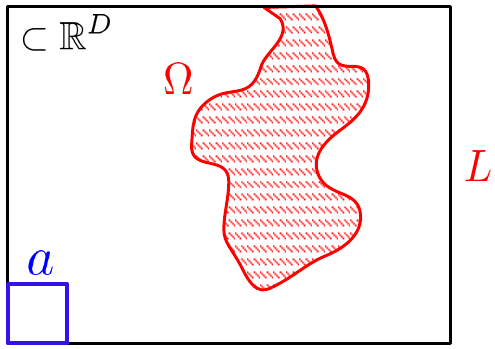}
	\vskip -0.00in      	
	\caption{\label{fig:IRdim} Left: Instead of the fixed subset $\Omega$ and 
	its measure, Effective-Number Theory shows how to define the effective 
	subset $\Omega_\text{eff}$ from the underlying probability distribution 
	$P(x)$. Right: IR dimension quantifies the change in the measure of 
	the set $\Omega$ (or $\Omega_\text{eff}$) in response to change of IR 
	cutoff $L$.}
	\vskip -0.16in
\end{figure}

\subsection{Non-Analyticity}
Phase transitions are characterized by non-analytic behavior of 
observables in thermodynamic limit. Identification of such 
non-analyticities in transitions to IR phase went hand-in-hand 
with formulation of the metal-to-critical 
picture~\cite{Alexandru:2021xoi, Alexandru:2021pap}. Indeed, 
dramatic feature of Anderson transitions is the spatial dimensional 
collapse of states on the localized side. Given the presence 
of Anderson-like mobility edges in IR phase, this aspect became
intuitively connected to roots of non-analyticity.
But formalization of dimensional arguments turned out to be a non-trivial 
task. Indeed, the notion of spatial dimension associated with Schr\"odinger
states (or with probability distributions in metric spaces more generally) 
didn't exist in the usual measure-based Hausdorff/Minkowski sense. 
This was rectified by the Effective-Number 
Theory~\cite{Horvath:2018aap} which constructed the needed effective 
counting measures (Fig.~\ref{fig:IRdim} left), and subsequently by 
the Effective-Dimension Theory~\cite{Horvath:2022ewv} which 
showed that these measures lead to the unique notion of effective 
dimension.

\begin{figure}[b]
	\centering  
	\vskip -0.19in      
	\includegraphics[width=0.78\linewidth]{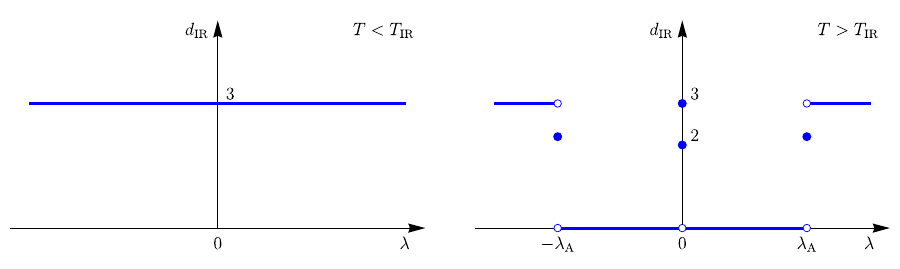}
	\vskip -0.05in      	
	\caption{\label{fig:dIR_spectral} IR dimensions of QCD Dirac modes in 
	the B phase (left) and in the IR phase (right)~\cite{Alexandru:2023xho}.}
	\vskip -0.16in
\end{figure}
    
Rather than UV dimension familiar from analyses of fractal sets, it is 
the unconventional IR dimension~\cite{Horvath:2022ewv} that is of prime 
interest here. While UV dimension gauges the response of set measure to 
the decrease of UV cutoff $a$, IR dimension probes the increase of 
IR cutoff~$L$, namely 
\begin{equation}
     N_+(\Omega, a, L) \propto L^{d_\fir(\Omega,a)}  
     \quad\quad \text{or} \quad\quad
     \efNm(P,a,L) \propto L^{d_\fir(P,a)} 
     \quad\quad \text{for} \quad\quad L \to \infty  \quad     
\end{equation}
for fixed sets and probability distributions respectively~\cite{Horvath:2022ewv}.
Here $N_+$ is the number of elementary volume elements covering $\Omega$ 
at given $a$ and $L$, while $\efNm$ is the minimal effective 
count~\cite{Horvath:2018aap} of these elements given the $P(x)$ in question.
Schematics of this is shown in the right plot of Fig.~\ref{fig:IRdim}.   

\begin{figure}[t]
	\centering  
	\vskip -0.20in      
	\includegraphics[width=0.70\linewidth]{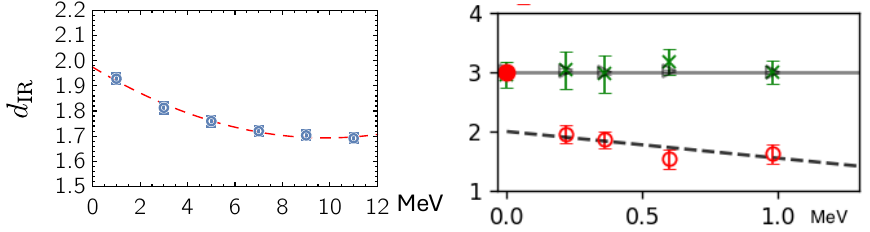}
	\vskip -0.04in      		
	\caption{\label{fig:deepIR_dim} IR dimensions of near-zero (critical)
	modes in IR phase. On the x-axis is the width of the near-zero region.  
	Left~\cite{Alexandru:2021pap}: pure-glue QCD at $T\!=\! 0.12 T_\fir$. 
	Right~\cite{Meng:2023nxf}: $N_f \!=\! 2+1$ real-world QCD 
	at $T \!=\! 234\,$MeV (red) and at $T \!=\! 187\,$MeV 
	(green, outside of IR phase). Full symbols are exact zeromodes.}
	\vskip -0.20in
\end{figure}

For Dirac modes of QCD and Schr\"odinger states of Anderson models,  
IR dimension is a function of Dirac scale ($\lambda$) or energy ($E$), and 
$P(x)$ is encoded by modes/states.
Effective counts $\efNm$ on which $d_\fir$ is based are QCD/disorder averaged. 
Extensive calculations~\cite{Alexandru:2021pap, Meng:2023nxf, Alexandru:2023xho}
support the spectral portrait of IR dimensions in QCD shown schematically in 
Fig.~\ref{fig:dIR_spectral}. While $d_\fir \!=\!3$ across the Dirac spectrum for 
$T \!<\! T_\fir$, in IR phase there is a pattern of discontinuities 
at mobility edges $\pm \lamA$ and $\lambda_\fir \!=\! 0$. Intriguing aspect
is that exact zeromodes have $d_\fir \!=\!3$, but the lowest near-zero (critical)
modes have $d_\fir \!=\!2$. The relevant evidence for this is shown in 
Fig.~\ref{fig:deepIR_dim} for both
the pure-glue~\cite{Alexandru:2021pap} and the $N_f \!=\! 2+1$ real-world 
QCD~\cite{Meng:2023nxf}. In the latter case the data nicely distinguishes 
the thermal state at $T \!=\! 187\,$MeV, which is not in IR phase, from 
that at $T \!=\! 234\,$MeV which is inside the phase.

\begin{figure}[t]
	\centering  
	\vskip -0.18in   
	\includegraphics[width=0.54\linewidth]{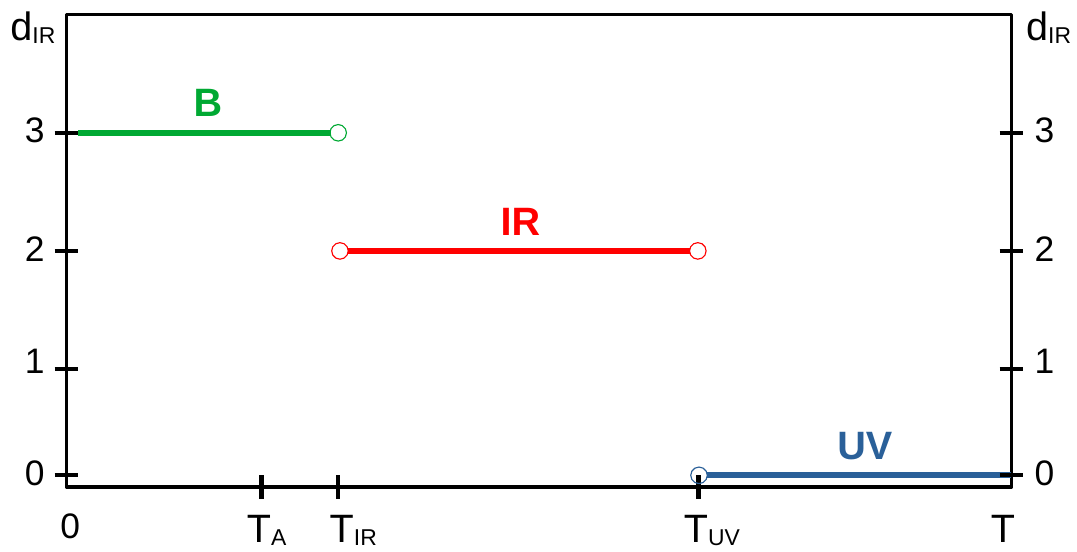}
	\vskip -0.00in      	
	\caption{\label{fig:dIR_Tdep} T-dependence of $d_\fir$ for deep-IR 
	Dirac modes which is also the spatial dimension of IR part~in~$F^2$.} 
	\vskip -0.14in
\end{figure}

Given the singular accumulation of near-zero modes in IR phase, $d_\fir \!=\!2$ 
dominates its deep-IR physics: non-analyticity of $\rho(\lambda)$ leads to
non-analyticity of observables at $T\!=\!T_\fir$. For example, the above 
results imply $d_\fir(T)$ for the IR part of action density shown in 
Fig.~\ref{fig:dIR_Tdep} 
(See Refs.~\cite{Horvath:2006az, Horvath:2006md, Alexandru:2008fu}.~). 

\section{Conclusions}

The chief purpose of this talk is to convey that, by virtue of the new 
IR phase, Anderson-like localization found a relevant place in 
the Standard Model of particle physics~\cite{Alexandru:2021xoi}. 
While a precise relation between ``Anderson and Anderson-like'' is yet 
to be determined, it is clear that the ideas developed in the former context 
are very fruitful for building a detailed understanding of the recently
uncovered IR regime represented by the IR 
phase.\footnote{$\,$I wish to acknowledge few additional
works ~\cite{Edwards:1999zm, Dick:2015twa, Kehr:2023wrs, Kovacs:2023vzi}  
related in various ways to the developments described here.} 

\acknowledgments
I greatfully acknowledge the long-term productive collaboration with Andrei
Alexandru and the extensive technical help from Dimitris Petrellis. Hali, Sylvia 
and Vlado made this text possible.  

\end{document}